Review article

# Optimization of laser illumination configuration for directly driven inertial confinement fusion

Masakatsu Murakami*, Daiki Nishi

*Institute of Laser Engineering, Osaka University, Osaka 565-0871, Japan*

## Abstract

Optimum laser configurations are presented to achieve high illumination uniformity with directly driven inertial confinement fusion targets. Assuming axisymmetric absorption pattern of individual laser beams, theoretical models are reviewed in terms of the number of laser beams, system imperfection, and laser beam patterns. Utilizing a self-organizing system of charged particles on a sphere, a simple numerical model is provided to give an optimal configuration for an arbitrary number of laser beams. As a result, such new configurations as "M48" and "M60" are found to show substantially higher illumination uniformity than any other existing direct drive systems. A new polar direct-drive scheme is proposed with the laser axes keeping off the target center, which can be applied to laser configurations designed for indirectly driven inertial fusion.
Copyright © 2016 Science and Technology Information Center, China Academy of Engineering Physics. Production and hosting by Elsevier B.V. This is an open access article under the CC BY-NC-ND license (http://creativecommons.org/licenses/by-nc-nd/4.0/).

*PACS Codes:* 28.52.Av; 52.57.Bc; 52.57.-z

*Keywords:* Analytical model; Laser illumination design; Polar direct drive; Inertial confinement fusion

## 1. Introduction

To achieve ignition and burn in inertial confinement fusion (ICF), a principal requirement is to illuminate a fusion pellet as uniformly as possible with a limited number of laser beams. The world's two largest laser fusion facilities, the US National Ignition Facility (NIF) [1] and the French Laser Mégajoule (LMJ) [2], are expected to achieve laboratory ignition based on the central spark ignition [3,4] in terms of the indirect-drive scheme, in which laser beams irradiate a high-Z casing surrounding a fuel pellet to produce thermal radiation [5]. This secondary thermal radiation can drive a spherically uniform implosion of the pellet, but at the price of low energy coupling. In contrast, in direct drive, as the counter scheme to the indirect-drive, laser beams directly illuminate a spherical pellet, also pursuing fast ignition driven by hot-electrons [6] and protons [7], impact ignition [8], or shock ignition [9] as well as the central spark ignition. In either case, uniform illumination of a fuel pellet is always a central issue in ICF and dominates in the target and reactor designs [10,11].

The implosion nonuniformity can be decomposed into factors of various harmonic modes. High-mode nonuniformities potentially drive Rayleigh–Taylor (R–T) instabilities. It is expected that the R–T instabilities are suppressed to a tolerable level by means of beam-smoothing techniques [12–15] and also physical mechanisms such as electron- and radiation-conduction [16,17]. Zooming techniques [18–20] are expected to significantly enhance time-integrated laser absorption in direct-drive, but can also introduce non-uniformity if not properly tailored to each implosion. Moreover, there is an important laser–plasma interaction referred to as cross-beam energy transfer (CBET) [21–23], which may significantly degrade laser absorption. In contrast to the high-mode nonuniformities, low-mode non-uniformities ($l \lesssim 10 - 20$) are hard to suppress even by the means mentioned above. To minimize the low-mode nonuniformities, optimization of laser illumination configuration is indispensable.

\* Corresponding author.
  *E-mail address:* murakami-m@ile.osaka-u.ac.jp (M. Murakami).

So far, various direct-drive irradiation systems have been designed for different numbers of laser beams $N_B$ [18,24–31], e.g., $N_B = 12$ (dodecahedron or Gekko XII [32]), 20 (icosahedron), 24 (modified rhombicuboctahedron or OMEGA [33]), 32 (20 face centers and 12 vertices of icosahedron) [34], and 60 (truncated icosahedron or OMEGA-upgrade [35]; hereafter referred as Ω60). All these designs are based on Platonic or Archimedean solids [36], which are applicable only to a specific set of numbers corresponding to the solids' shape. Here a new numerical scheme is presented to give an optimum direct-drive beam configuration for an arbitrary number of beams, which is obtained as a self-organizing system by solving an $N$-body charged particle simulation [37]. As a result, for example, a new forty-eight-beam system "M48" (the detail is given below) has been found to show substantially higher illumination uniformity than any other existing direct drive systems based on the ancient geometry.

In this paper, we review some theoretical work for optimizing beam configurations for direct-drive illumination as well as introducing a new scheme for polar direct drive (PDD). Note that an excellent review on direct-drive inertial confinement fusion can be found in Ref. [38].

The basic assumption employed throughout the present paper is that the physical quantities of the target and surrounding corona plasma have all spherically symmetric profiles and that a single laser beam has axially symmetric intensity profile with respect to its beam axis. It follows that, in particular for the PDD design given here, the analysis better applies to the early stage of implosion, when the target deformation from sphericity can still be neglected. Besides, it is assumed that the corona layer surrounding a spherical target, in which laser rays propagate and deposit their energies, are infinitesimally thin. Although the present paper contains such many assumptions, it provides a reasonable set of results including many features that would benefit a full radiation-hydrodynamic simulation.

The structure of this paper is as follows: In Sec. 2, an analytical model is given to evaluate the root-mean-square (rms) non-uniformity assuming that every single laser beam has a similar axisymmetric beam pattern but can have different energy. Based on the axisymmetric assumption, one obtains a useful function as the geometrical factor [33] that one can assess the illumination performance from pure geometrical point of view, which is crucial when optimizing an illumination configuration. In Sec. 3, optimum configurations are determined for axisymmetric irradiation system, in which all the laser beam axes go through the target center and are arranged in a cone shape. In Sec. 4, a new analytical model for PDD is proposed, in which the laser beams are kept off the target center. It should be noted that the present work goes beyond the previous works on PDD [30,39], which were obtained rather empirically, on the point of determining optimum configurations mechanically as an eigenvalue problem. In Sec. 5, a numerical model is presented to determine a best configuration for an arbitrary number of laser beams based on a self-organizing system of charged particles. In terms of the analysis, it is shown that more laser beams does not necessarily bring about higher illumination uniformity. Finally Sec. 6 is devoted for a summary.

## 2. Analytical model for uniformity evaluation

### 2.1. Geometrical factor

In this subsection, we briefly review an analytical model by Skupsky and Lee [33] and in the next subsection we extend it to evaluate illumination uniformity under system imperfections [40]. We assume that the absorbed pattern of every single beam has a similar axisymmetric profile to each other, but each beam can have different amplitudes. This axisymmetric assumption can be used only if the beam axes pass through the target center, therefore it applies to Secs. 3 and 5 but does not apply to the PDD analysis given in Sec. 4, where the beam axes are offset from the target center. The absorbed intensity profile of the $k$th laser beam is expanded in Legendre polynomials,

$$I_k(\gamma) = \overline{I_k}\left[1 + \sum_{l=1}^{\infty} a_l P_l(\cos\gamma)\right], \quad (1)$$

where $\overline{I_k}$ denotes the average intensity over the sphere, and the $l$th coefficient,

$$a_l = \frac{2l+1}{2} \int_{-1}^{1} \frac{I_k(\gamma)}{\overline{I_k}} P_l(\cos\gamma) \mathrm{d}(\cos\gamma), \quad (2)$$

is common for each beam. The angle $\gamma$ is taken between the beam axis $\widehat{\Omega}_k$ (the hat denotes a unit vector), and an observing point $\widehat{r}$, i.e., $\cos\gamma = \widehat{r} \cdot \widehat{\Omega}_k$. The total absorbed intensity is the sum of Eq. (1) over all beams:

$$I_a(\widehat{r}) = I_T + \sum_{l=1}^{\infty} a_l \sum_{k=1}^{N_B} \overline{I_k} P_l(\widehat{r} \cdot \widehat{\Omega}_k), \quad (3)$$

where $I_T = \sum_k \overline{I_k}$. The rms deviation is then defined by

$$\sigma_{\mathrm{rms}} = \left\{\frac{1}{4\pi I_T^2} \int [I_a(\widehat{r}) - I_T]^2 \mathrm{d}\widehat{r}\right\}^{1/2} = \left(\sum_{l=1}^{\infty} \sigma_l^2\right)^{1/2}, \quad (4)$$

where $\sigma_l$ is the $l$th component obtained by the use of orthogonal property of spherical harmonics, which is further decomposed as a product in the form,

$$\sigma_l = \frac{a_l}{\sqrt{2l+1}} G_l, \quad (5)$$

where the first term $a_l/\sqrt{2l+1}$ is the single beam factor, while the second term is the geometrical factor defined by

$$G_l = \left[\sum_{j=1}^{N_B}\sum_{k=1}^{N_B} P_l(\widehat{\boldsymbol{\Omega}}_j \cdot \widehat{\boldsymbol{\Omega}}_k)\overline{I}_j\overline{I}_k/I_T^2\right]^{1/2}. \quad (6)$$

As can be seen from Eq. (6), the geometrical factor $G_l$ includes all the basic information, i.e., the pointing and power of an individual beam, and therefore $G_l$ provides very essential geometrical spectrum of a given illumination system.

The formalism of Eqs. (1)−(6), first proposed by Skupsky and Lee [33], treats a specific spherical surface or approximately a very thin layer compared to the radius. Therefore the model cannot rigorously describe such a physical picture as laser absorption over a long range in the plasma and resultant three-dimensional (3D) absorption pattern. However, as long as the 3D absorption distribution of an individual beam is still axisymmetric and overlapping with the others can be treated as linear, the above formalism is always valid when assessing direct-drive laser configuration. In other words, even at different depth in the radial direction characterized by different single beam spectrum $a_l$, the spectrum of $G_l$ has its own eigenstructure and is kept unchanged once the laser illumination configuration is fixed.

Fig. 1 shows $G_l$ for different number of beams $N_B$ and mode number $l$. The label "P32" denotes the solid body with 32 vectors corresponding to 12 face centers and 20 vertices of one of the Platonic solids, dodecahedron, while "M48", "M60", and "M72" denote other specific configurations with 48, 60, and 72 vectors that are obtained by a new algorithm utilizing self-organizing system [37] as will be discussed in Sec. 4. The advantages of increasing $N_B$ can be easily seen in Fig. 1. By increasing $N_B$, the amplitudes of $G_l$ in general decrease, and the lowest dominant mode $n_d$ increases, which is approximately given by $n_d \sim \pi\sqrt{N_B}/2$ [40]. Note that the geometrical spectra for P32 and $\Omega 60$ are similar to each other in spite that the numbers of beams are different by a factor of two. Their common lowest dominant mode is $n_d = 10$. The geometrical factor $G_l$ also plays an important role when evaluating hohlraum radiation uniformity [41]. Recently, Lan et al. proposed hexahedral hohlraum target [42,43]. In their hohlraum design, the lowest dominant mode is $n_d = 4$, which is at a glance substantially lower than with direct-drive configuration, however in indirect-drive targets, the uniformities are strongly smeared out by X-ray radiation [41,44].

### 2.2. Imperfection effects

Below we find how system imperfections degrade the illumination uniformity [40] by assessing the behavior of $G_l$. We first consider power imbalance effect, where the pointing of the beams is assumed to be perfect, but the beam powers are fluctuated by the factors $\varepsilon_k$ as follows,

$$\overline{I}_k = I_0(1+\varepsilon_k), \quad \sum_{k=1}^{N_B}\varepsilon_k = 0, \quad \left(\sum_{k=1}^{N_B}\varepsilon_k^2/N_B\right)^{1/2} = \sigma_P, \quad (7)$$

where $I_0 = I_T/N_B$ is the average intensity and $\sigma_P$ represents the rms deviation of power imbalance. With the help of Eq. (7), the geometrical factor defined by Eq. (6) is then rewritten in the form

$$G_l = \left[(G_l^0)^2 + (\Delta G_l)^2\right]^{1/2}, \quad (8)$$

where

$$G_l^0 = \left[\sum_{j=1}^{N_B}\sum_{k=1}^{N_B} P_l(\widehat{\boldsymbol{\Omega}}_j^0 \cdot \widehat{\boldsymbol{\Omega}}_k^0)/N_B^2\right]^{1/2}, \quad (9)$$

$$\Delta G_l = \left[\sum_{j=1}^{N_B}\sum_{k=1}^{N_B} P_l(\widehat{\boldsymbol{\Omega}}_j^0 \cdot \widehat{\boldsymbol{\Omega}}_k^0)\varepsilon_j\varepsilon_k/N_B^2\right]^{1/2}. \quad (10)$$

The average $\Delta G_l^{ave}$ and the rms deviation $\Delta G_l^{rms}$ of $\Delta G_l$ are of high importance and interest, which are respectively given by

$$\Delta G_l^{ave} = \frac{\sigma_P}{\sqrt{N_B}}, \quad \Delta G_l^{rms} = \frac{\sigma_P}{\sqrt{2N_B}}. \quad (11)$$

Next, we assess the pointing error effect assuming perfect power balance among the beams, but the beam pointing (the axes direction $\widehat{\boldsymbol{\Omega}}_k$) are deteriorated. The geometrical factor is then rewritten by

$$G_l = \left\{\sum_{j=1}^{N_B}\sum_{k=1}^{N_B} P_l\left[\left(\widehat{\boldsymbol{\Omega}}_j^0 + \Delta\widehat{\boldsymbol{\Omega}}_j\right)\cdot\left(\widehat{\boldsymbol{\Omega}}_k^0 + \Delta\widehat{\boldsymbol{\Omega}}_k\right)\right]/N_B^2\right\}^{1/2}, \quad (12)$$

where the perturbed beam axes are still unit vectors, i.e., $\left|\widehat{\boldsymbol{\Omega}}_k^0 + \Delta\widehat{\boldsymbol{\Omega}}_k\right| = 1$ is kept for all $k$'s. Assuming that $\Delta\widehat{\boldsymbol{\Omega}}_k$ has Maxwellian distribution with its rms deviation, $\sigma_\Omega$, given in the form,

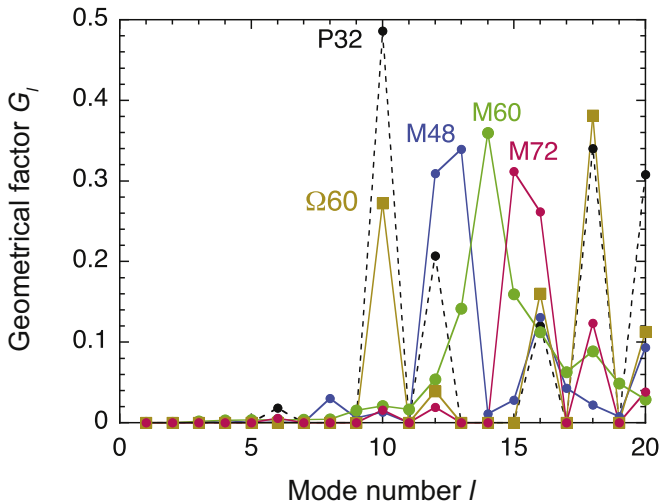

Fig. 1. Geometrical factors for different configurations with $N_B = 32-72$. A good measure of an illumination configuration is "how many lower modes are suppressed in the spectrum of the geometrical factor". Based on the measure, the superiority of the direct drive illumination configurations is read to be M72 > M60 > M48 > $\Omega 60$ > P32 in this figure.

$$\sigma_\Omega = \left[\sum_{k=1}^{N_B}(\Delta\widehat{\boldsymbol{\Omega}}_k)^2/N_B\right]^{1/2}, \tag{13}$$

the statistical average of $G_l$ is given by

$$\Delta G_l^{\text{ave}} = \begin{cases} l\sigma_\Omega/\sqrt{2N_B}, & l\sigma_\Omega \lesssim 1, \\ 1/\sqrt{N_B}, & l\sigma_\Omega \gtrsim 1. \end{cases} \tag{14}$$

Thus the averages $\Delta G_l^{\text{ave}}$, subject to power imbalance and pointing error, have the same dependency in proportion to $1/\sqrt{N_B}$.

When the imperfections $\sigma_P$ and $\sigma_\Omega$ are both included in the irradiation system, one can obtain a simple formula for $\sigma_{\text{rms}}$ in terms of the canonical one $\sigma_{\text{rms}}^0$ ($\sigma_P = \sigma_\Omega = 0$) and the deterioration $\Delta\sigma_{\text{rms}}$ with the aid of Eqs. (4), (5), (8), (10) and (14),

$$\sigma_{\text{rms}} = \sqrt{(\sigma_{\text{rms}}^0)^2 + (\Delta\sigma_{\text{rms}})^2}, \tag{15}$$

where

$$\Delta\sigma_{\text{rms}} = \sigma_{\text{rms}}^{\text{single}} \frac{\sigma_{\text{sys}}}{\sqrt{N_B}}, \tag{16}$$

is further composed of the rms deviation of the single beam,

$$\sigma_{\text{rms}}^{\text{single}} = \left(\sum_{l=1}^\infty \frac{a_l^2}{2l+1}\right)^{1/2}, \tag{17}$$

and the system imperfection $\sigma_{\text{sys}}$ given by

$$\sigma_{\text{sys}} = \sqrt{\sigma_P^2 + (\bar{l}\sigma_\Omega)^2}, \tag{18}$$

where

$$\bar{l} = \frac{1}{\sigma_{\text{rms}}^{\text{single}}}\left[\sum_{l=1}^\infty \frac{l^2 a_l^2}{2(2l+1)}\right]^{1/2}, \tag{19}$$

is the effective mode number. Both $\sigma_{\text{rms}}^{\text{single}}$ and $\bar{l}$ depend on the pattern of an individual beam, and are in general limited in a narrow region between 1 and 3, if the single beam pattern has a smooth curve. If $\sigma_{\text{rms}}^0$ is small enough (after all we have to seek such a combination of a single beam pattern and a configuration of beams to satisfy $\sigma_{\text{rms}}^0 \leq 0.1\% - 0.5\%$), then $\sigma_{\text{rms}} \approx \Delta\sigma_{\text{rms}}$ practically holds for values of $\sigma_{\text{sys}}$ on the order of 1%. In practical systems, there are even more statistical fluctuations such as beam-to-beam shapes and mistiming. Such errors can be taken into account approximately in Eq. (18) in the form as $\sigma_{\text{sys}} = \sqrt{\sigma_1^2 + \sigma_2^2 + \sigma_3^2 + ...}$, as long as the errors are small enough and statistically irrelevant to each other.

Here we demonstrate that the model with axisymmetric assumption can also be applied to axially asymmetric patterns. Suppose that asymmetric $k$th beam pattern is approximately decomposed into two axisymmetric patterns, whose beam axes are $\widehat{\boldsymbol{\Omega}}_k^0$ and $\widehat{\boldsymbol{\Omega}}_k^0 + \Delta\widehat{\boldsymbol{\Omega}}_k$, in the form,

$$I_k(\widehat{\boldsymbol{r}}) = I_0\left(1 + \sum_{l=1}^\infty\left\{a_l P_l\left(\widehat{\boldsymbol{r}}\cdot\widehat{\boldsymbol{\Omega}}_k^0\right) + b_l P_l\left[\widehat{\boldsymbol{r}}\cdot\left(\widehat{\boldsymbol{\Omega}}_k^0 + \Delta\widehat{\boldsymbol{\Omega}}_k\right)\right]\right\}\right). \tag{20}$$

Then, using the orthogonal property of spherical harmonics, the rms deviation is worked out to be

$$\sigma_{\text{rms}} \approx \left\{\sum_{l=1}^\infty\left[\frac{(a_l+b_l)^2}{2l+1}(G_l^0)^2 + \frac{b_l^2}{2l+1}\frac{l(l+1)\sigma_\Omega^2}{2N_B}\right]\right\}^{1/2}, \tag{21}$$

where $\sigma_\Omega \ll 1$ is again assumed. Since the first term can often be negligible for lower modes in an optimized system, Eq. (21) is further reduced to $\sigma_{\text{rms}} \approx \sigma_{\text{rms}}^{\text{single}}\bar{l}\sigma_\Omega/\sqrt{N_B}$, where $\sigma_{\text{rms}}^{\text{single}}$ is relevant to the coefficients $b_l$. Thus the nonuniformity with asymmetric absorption patterns can be equivalently evaluated by an effective imperfection $\sigma_\Omega$. For more general cases, one can easily extend the expression of Eq. (20) by substituting the asymmetric part for $\Sigma_i b_l P_l[\widehat{\boldsymbol{r}}\cdot(\widehat{\boldsymbol{\Omega}}_k^0 + \Delta\boldsymbol{\Omega}_k^i)]$.

In this subsection we did not address "mis-pointing" (off-center illumination keeping the original pointing $\widehat{\boldsymbol{\Omega}}_k$), which is essentially different from the "pointing error" described by Eqs. (12)−(14). This mis-pointing is much more likely to occur in practical systems. However, the mis-pointing can also be treated by combining the analyses used for the "pointing error" and the above asymmetric pattern (the detailed analysis is not given in this paper).

## 3. Optimization of multi-cone configuration

The conic irradiation system presented here provides a mechanical scheme, with which $\sigma_{\text{rms}}$ can be reduced with increasing $N_B$ [46]. Fig. 2 shows the schematic picture of a conic irradiation system. The beam axes pointing to the target center are arranged on $N_C$ cone surfaces ($N_C = 4$ in Fig. 2); the $i$th cone is characterized by the polar angle $\Theta_i$, and the total energy $E_i$. For simplicity, we start with the limiting case where the number of beams is infinitely large ($N_B = \infty$) and are divided into $N_C$ cones characterized by cone angles $\Theta_1, \Theta_2, \cdots, \Theta_{N_C}$, and energies $E_1, E_2, \cdots, E_{N_C}$. It should be noted that the ratios of the beam numbers between the groups do not matter for the moment. In the limit of $N_B = \infty$, the geometrical factor, Eq. (6), is rewritten in the form,

$$G_l = \left[\sum_{i=1}^{N_C}\sum_{j=1}^{N_C}E_i E_j \int_0^\pi P_l(\cos\gamma_{ij})d\phi \middle/ \pi\sum_{i=1}^{N_C}\sum_{j=1}^{N_C}E_i E_j\right]^{1/2}, \tag{22}$$

where

$$\cos\gamma_{ij} = \sin\Theta_i\sin\Theta_j\cos\phi + \cos\Theta_i\cos\Theta_j. \tag{23}$$

With the help of the formula,

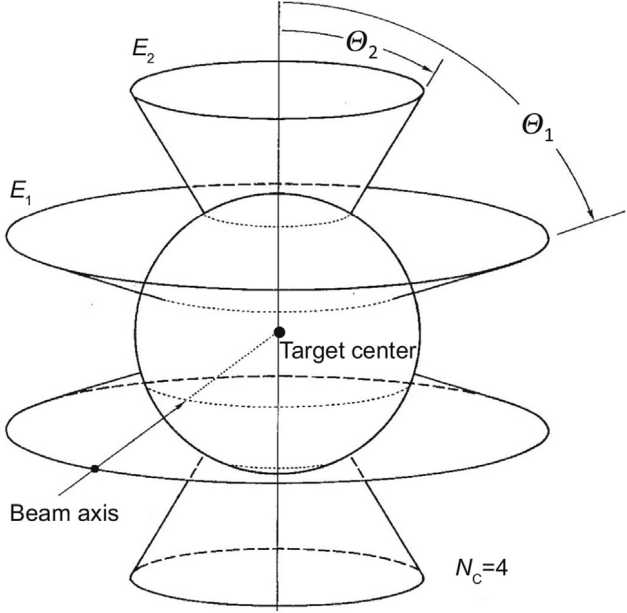

Fig. 2. Schematic view of the conic irradiation system for $N_C = 4$ case. The beam axes pointing to the target center are arranged on the cone surfaces.

$$\int_0^\pi P_l(\cos\gamma_{ij}) d\phi = \pi P_l(\cos\Theta_i) P_l(\cos\Theta_j), \quad (24)$$

then Eq. (22) reduces to a simple form,

$$G_l = \sum_{i=1}^{N_C} E_i P_l(\cos\theta_i) \Big/ \sum_{i=1}^{N_C} E_i. \quad (25)$$

To eliminate the non-uniformities of all odd modes, the beams must be arranged symmetrically with respect to the target center. Apparently, this implies that the cone configurations must be symmetric with respect to the equator plane ($\theta = \pi/2$). For this reason, we are interested in only one hemisphere. The $N_C$-cone system has then $N_C - 1$ independent parameters, i.e. $N_C/2$ angles and $N_C/2 - 1$ relative energies. Mathematically this means that nonuniformities of $N_C - 1$ even modes can be intentionally smeared out regardless of the single beam pattern, by optimizing the parameters such that

$$G_2 = G_4 = \ldots = G_{2(N_C-1)} = 0. \quad (26)$$

Thus, the lowest dominant mode number is given by $n_d = 2N_C$. The $m$th equation in Eq. (26), $G_{2m} = 0$, is rewritten from Eq. (25) in the form,

$$\sum_{i=1}^{N_C/2} E_i P_{2m}(\cos\Theta_i) = 0. \quad (27)$$

Because the number of the independent parameters is equal to that of the simultaneous equations, we can expect nontrivial solutions to them. Table 1 gives such solutions for $\Theta_i$ and $E_i$ that are numerically obtained.

So far, we treated an idealized case of an infinitely large number of beams ($N_B = \infty$). However, $N_B$ is, of course, finite in real systems. Then, the first question would be: whether the optimum conditions obtained for $N_B = \infty$ are still valid for a finite number of beams, and if so, what is the minimum required number of $N_B$ for each $N_C$, and how many beams are to be assigned to each cone? To answer these questions, we have performed numerical calculations on the geometrical factor as Eq. (6). Thereby, the cone angles $\Theta_i$ and the energies $E_i$ are taken from Table 1. Suppose that a finite number of beams, $M_i$, is assigned to the $i$th cone ($i = 1, 2, \ldots, N_C$) under fixed number of $N_B = \Sigma_i M_i$. The beams are assumed to be uniformly distributed along the azimuthal direction on each cone surface, and have equal energy, $E_i/M_i$, for the $i$th cone. Furthermore, the rotations $\Delta\Phi_i$ of beam axes along the azimuthal direction are also taken into account. The azimuthal coordinate of $j$th beam of $i$th cone is then given by $\Phi_{ij} = 2\pi j/M_i + \Delta\Phi_i$ ($j = 1, 2, \ldots, M_i$). The possible combinations $\{M_1, M_2, \ldots, M_{N_C}\}$ are numerically found, and summarized in Table 2. It should be noted that the rotation $\Delta\Phi_i$ has turned out not to affect the numerical results at all. Also, for example, the case of hexahedron is classified into the double-cone configuration: $N_C = 2$, $M_1 = 3$, and $N_B = 6$.

Table 1
Optimum irradiation configuration for axisymmetric system (see Fig. 2 for $N_C = 4$ case). The polar angles, $\Theta_1 - \Theta_4$, are measured in the unit of degree. The energy assigned to the 1st cone $E_1$ is normalized to unity.

| $N_C$ | 1st cone | | 2nd cone | | 3rd cone | | 4th cone | |
|---|---|---|---|---|---|---|---|---|
| | $\Theta_1$ | $E_1$ | $\Theta_2$ | $E_2$ | $\Theta_3$ | $E_3$ | $\Theta_4$ | $E_4$ |
| 2 | 54.736 | 1 | | | | | | |
| 4 | 70.124 | 1 | 30.556 | 0.53340 | | | | |
| 6 | 76.195 | 1 | 48.608 | 0.77100 | 21.177 | 0.36615 | | |
| 8 | 79.430 | 1 | 58.296 | 0.86496 | 37.187 | 0.61315 | 16.201 | 0.27911 |

Table 2
Optimized configuration of number of beams on each cone (see Fig. 2); the corresponding polar angles and energies are given in Table 1. The beam number configurations given here give rough instruction on how many beams are necessary to keep the performance of $N_C$-cone design, i.e., $n_d = 2N_C$.

| $N_C$ | $n_d$ | $M_1$ | $M_2$ | $M_3$ | $M_4$ | $N_B$ |
|---|---|---|---|---|---|---|
| 2 | 4 | 3 | | | | 6 |
| 4 | 8 | 7 | 7 | | | 28 |
| 6 | 12 | 11 | 11 | 8 | | 60 |
| 8 | 16 | 15 | 15 | 13 | 8 | 102 |

## 4. Design of polar direct drive

In the previous Section, we treated axisymmetric system, where all the beam axes go through the target center. If the beam axes are allowed to be off target center, the illumination design becomes more flexible at least in the parametric space. In this Section, the absorbed laser patterns are not axisymmetric anymore as a result of the oblique incidence, while the cross-sectional pattern of an incident single beam is kept axisymmetric as before just for simplicity.

Here suppose a laser beam with an incident angle $\psi$ ($0 \leq \psi \leq \pi/2$) to the polar axis (see Fig. 3) characterized by the unit vector $\widehat{\Omega} = (\sin\psi, 0, \cos\psi)$. Moreover the beam axis

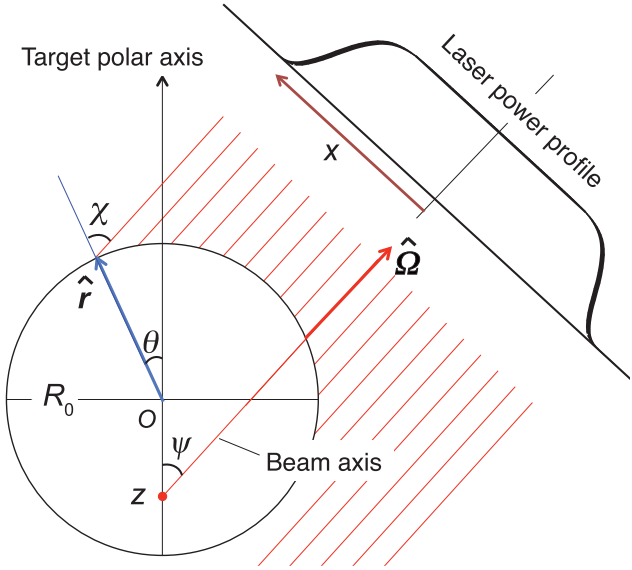

Fig. 3. Illumination configuration for polar direct drive.

The resultant absorbed intensity distribution of a single beam is then given as a function of $\chi(\theta, \phi, \psi)$ and $x(\theta, \phi, \psi, z)$ in the form,

$$I_a(\theta, \phi) = \eta_a(\chi) I_L(x) \cos\chi. \quad (32)$$

The area being illuminated by a single beam is given by

$$|\phi| \leq \begin{cases} \pi, & 0 \leq \theta \leq \frac{\pi}{2} - \psi; \\ \arccos(-\cot\psi\cot\theta), & \frac{\pi}{2} - \psi < \theta \leq \frac{\pi}{2} + \psi. \end{cases} \quad (33)$$

Fig. 4 shows the absorption pattern by a single beam for different values of $\psi$ and $z$ drawn on the $\theta-\phi$ plane. Changing the stand-off distance $z$ from $z = 0.5$ to $z = -0.5$ under the same pointing angle $\psi$, the absorption pattern moves to the right as could be expected. The same happens when increasing the angle $\psi$ from $\psi = \pi/6$ to $\psi = \pi/3$, keeping $z$ unchanged. Thus the absorption pattern can be controlled in two dimensions in terms of $\psi$ and $z$.

Suppose that a number of laser beams, which belong to the same "illuminating ring" and are characterized by a set of $\psi$ and $z$, illuminate a spherical target with the common absorption pattern and uniformly surround the target polar axis. The averaged absorption pattern $I_{av}$ is then given as a function of only $\theta$ in the form,

$$I_{av}(\theta) = \frac{1}{2\pi} \int_{-\pi}^{\pi} I_a(\theta, \phi) d\phi. \quad (34)$$

Fig. 5 shows the numerical result of $I_{av}(\theta)$ obtained for different $z$-values under the fixed condition, $\psi = \pi/3$, $\alpha = 0.6$, $\beta = 6$, and $\eta_\perp = 0.95$ (compare with Eqs. (29) and (30)). The two curves with $z = 0.5$ and $-0.5$ correspond to the upper-right and lower-right sub-figures in Fig. 4, respectively. The peak values are normalized to unity for simplicity.

The ring pattern $I_{av}(\theta)$ can be now expanded into Legendre polynomials,

$$I_{av}(\theta) = \bar{I}_{av}\left[1 + \sum_{l=1}^{\infty} b_l P_l(\cos\theta)\right], \quad (35)$$

where $\bar{I}_{av}$ denotes the average intensity over the sphere, and the coefficient $b_l$ is given by

$$b_l = \frac{2l+1}{2} \int_{-1}^{1} \frac{I_{av}(\cos\theta)}{\bar{I}_{av}} P_l(\cos\theta) d(\cos\theta). \quad (36)$$

Fig. 6 shows the amplitude $b_l$ obtained as a function of the incident angle of beam axis $\psi$ and the stand-off distance $z$ for different Legendre modes, $l = 2, 4, 6,$ and $8$. Apparently, with increase in $l$, the pattern of $b_l$ becomes ruffled more frequently on the $\psi-z$ plane. Such a behavior brings about multiple sets of eigenvalue solutions of the system, from which the most practical set for ICF implosions needs to be found.

When we consider $N_C$ cone system as was treated in the previous Section, we have $N_C/2$ cones on one hemisphere, and each cone has three independent parameters, i.e., the incident is assumed to intersect with the polar axis at coordinate $z$. The incident angle $\chi$ of a laser light hitting at an arbitrary point $\hat{r}(\theta, \phi)$ is given by

$$\cos\chi = \hat{r} \cdot \hat{\Omega} = \sin\theta\cos\phi\sin\psi + \cos\theta\cos\psi. \quad (28)$$

In the following, we assume that the spherical corona layer is thin enough and that all the incident laser rays are parallel to the beam axis. In the corona layer, the laser energies are absorbed along the trajectories and finally refracted out of the target sphere. This thin layer assumption is valid only for the early stage of implosion. The corona plasma is approximated to have an exponentially decaying density profile along the radius with a scale length $L = |n_c(\partial n_e/\partial r)^{-1}|$, where $n_c$ is the critical density, and the laser rays undergo collisional absorption along their trajectories. Then the absorption efficiency of a laser ray with an incident angle $\chi$ is computed by [47]

$$\eta_a(\chi) = 1 - (1 - \eta_\perp)^{\cos^3\chi}, \quad (29)$$

where $\eta_\perp = 1 - \exp(-8\nu_{ei}L/3c)$ is the absorption efficiency of a normally incident laser ($\chi = 0$), $\nu_{ei}$ is the electron-ion collision frequency at the critical density; $c$ is the speed of light. Furthermore, we employ a super-Gaussian intensity profile as the single beam profile in the form,

$$I_L(x) = I_0 \exp\left[-\left(\frac{x}{\alpha R_0}\right)^\beta\right], \quad (30)$$

where $I_0$ is the laser intensity on the beam axis, $\alpha$ and $\beta$ are the control parameters of the beam shape, $x$ is the distance from the beam axis, given under the present configuration by

$$x = R_0\sqrt{(\sin\theta\sin\phi)^2 + (z - \cos\theta + \sin\theta\cos\phi\cot\psi)^2 \sin^2\psi}. \quad (31)$$

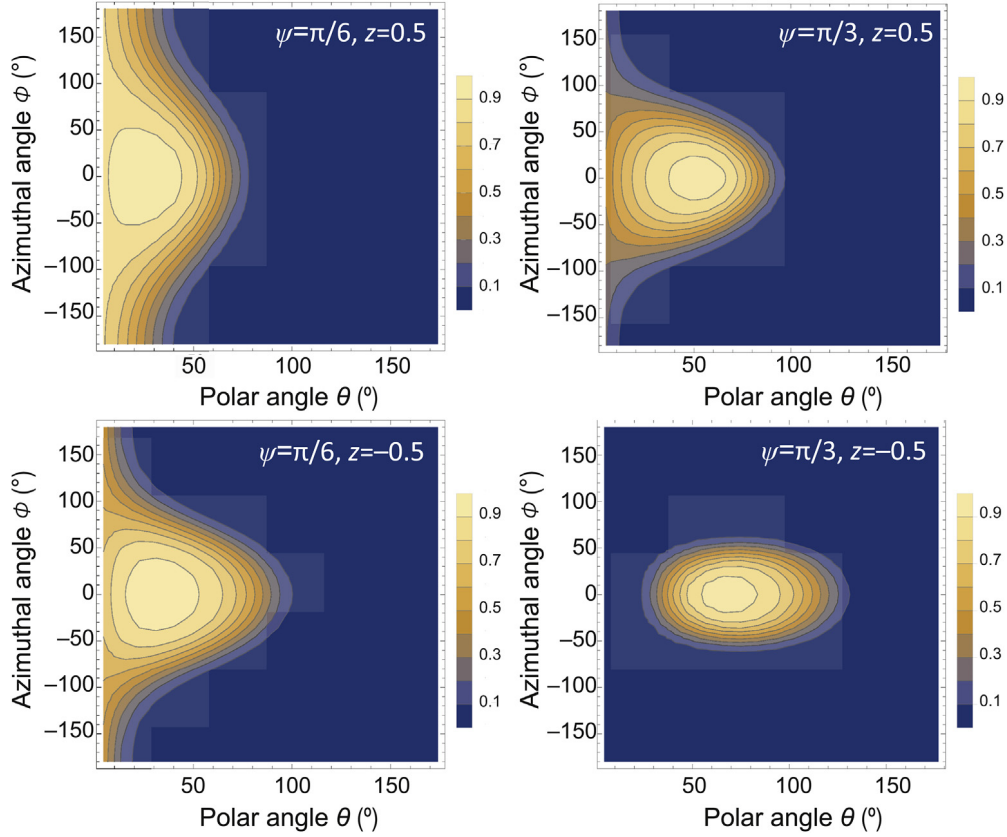

Fig. 4. Absorption pattern illuminated by a single beam for different values of $\psi$ and $z$ on the $\theta-\phi$ plane. The horizontal and vertical axes stand for the polar angle $\theta$ and the azimuthal angle $\phi$ in degrees, respectively. Fixed parameters are $\alpha = 0.9$, $\beta = 6$, $\eta_\perp = 0.95$.

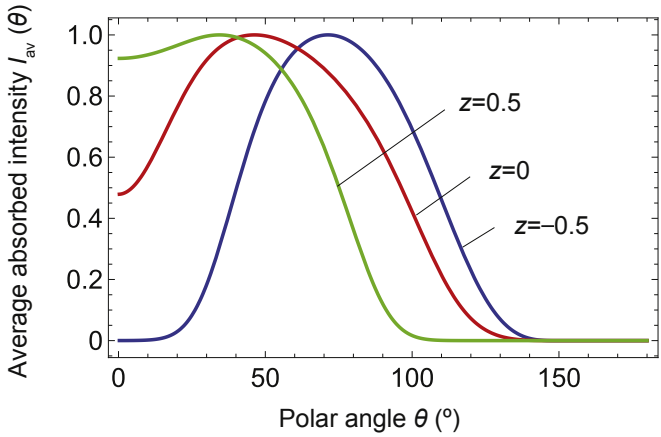

Fig. 5. Averaged absorbed intensity $I_{av}(\theta)$. Fixed parameters are $\alpha = 0.9$, $\beta = 6$, $\eta_\perp = 0.95$, $\psi = \pi/3$, while $z$ is varied for three different values.

angle $\psi$, the stand-off distance $z$, and the total energy $E$. Since the energies can be normalized to one of a specified cone (for example $E_1 = 1$ as in Table 1), one has $3N_C/2-1$ free parameters after all, in other words, one can intentionally delete $3N_C/2-1$ lowest modes by optimizing the parameters. This can be expressed such that the total amplitude $B_{2m}$ of $2m$th Legendre mode integrated over $N_C$ cones can be smeared out by satisfying the following equations,

$$B_{2m} = \sum_{i=1}^{N_C/2} E_i b_{2m}(\psi_i, z_i) = 0, \quad m = 1, 2, \ldots, \frac{3}{2}N_C - 1. \quad (37)$$

This relation corresponds to Eq. (26) for the axisymmetric design with all the beam axes passing through the target center. In case of $N_C = 4$ corresponding to Fig. 2, five unknown parameters, $\psi_1$, $\psi_2$, $z_1$, $z_2$, and $E_2$ are numerically optimized to achieve $B_2 = B_4 = \ldots = B_{10} = 0$ by solving the simultaneous equations Eq. (37) to give a set of solution, $\psi_1 = 46°$, $\psi_2 = 53°$, $z_1 = -0.65$, $z_2 = 0.80$, $E_1 = 1$, and $E_2 = 0.40$ for the fixed parameters, $\alpha = 0.6$, $\beta = 6.0$, $\eta_\perp = 0.9$ (it gives just one of the several solutions).

The set of the parameters can be compared with the optimal design for normally directed system given in Table 1 ($N_C = 4$), e.g., $\theta_1 = 70°$, $\theta_2 = 30°$, $E_1 = 1$, and $E_2 = 0.53$, which are obtained regardless of the values of $\alpha$, $\beta$, and $\eta_\perp$ because the beam pointing and single beam pattern are separated from each other under the condition, $z_1 = z_2 = 0$. In the same manner, one can obtain optimum illumination configurations for an arbitrary number of $N_C$. The numerical scheme given here can also be applied to such systems as NIF and LMJ, which already have fixed incident angles $\psi_i$, and therefore the stand-off distances $z_i$ and assigned energies $E_i$ are the parameters to be optimized. More detailed analysis, taking the temporal behaviors of target implosion and pulse shape into account, is now under way.

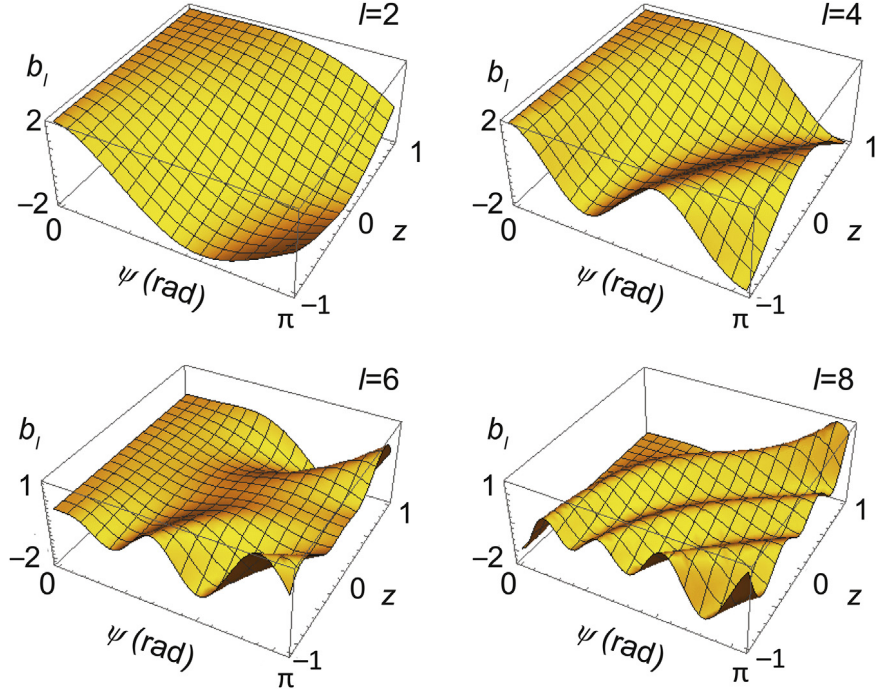

Fig. 6. Amplitude $b_l$ for laser absorption with respect to a single cone configuration, which is obtained as a function of the incident angle of beam axis, $\psi$, and the stand-off distance, $z$ (referred to Fig. 3). The parameters, $\alpha = 0.9$, $\beta = 6$, $\eta_\perp = 0.9$, are fixed.

## 5. Self-organizing Coulomb system to optimize beam pointing

Here we describe a totally new approach to find optimum beam configurations in ICF, which applies to direct drive scheme with all the beam axes passing through the target center. Suppose that $N_B$ charged particles are randomly distributed on a sphere at $t = 0$ and are dynamically redistributed due to the Coulomb repulsion to settle at a fixed configuration in a self-organized manner [37]. Our hypothesis is then as follows: If one finds such a configuration of $N_B$ charged particles on a sphere that has the lowest Coulomb potential energy, the resulting system gives the highest irradiation uniformity. The radial beam axes are determined from the particle positions obtained. The present method reproduces Platonic solids with $N_B = 4, 6, 8$, and $12$ except for $N_B = 20$. Moreover, such new configurations as M48 and M72 have turned out to demonstrate even better performance in achieving higher irradiation uniformity than the other configurations, where the notation "M48", for example, stands for the solution obtained by the present method with $N_B = 48$. In addition, we use the notation "P12" (dodecahedron), "P20" (icosahedron), and "P32" (12 faces and 20 vertices of dodecahedron) for Platonic solids and their secondary solids. Here it should be noted that Refs. [45,46] propose cylindrically symmetric irradiation systems with different number of beams, which are substantially different from spherically symmetric configurations obtained by the present method. In fact, according to numerical surveys (details of the comparisons are not presented in this paper), the latter (spherical) shows higher performance as an illumination system than the former (cylindrical).

The equation of motion for the $i$th charged particle ($1 \leq i \leq N_B$), in which all the physical quantities are appropriately normalized (unit mass, unit charge, etc.) and thus becomes dimensionless, is given by

$$\frac{d^2 \widehat{r}_i}{dt^2} = \sum_{j=1 (j \neq i)}^{N_B} \frac{\widehat{r}_i - \widehat{r}_j}{|\widehat{r}_i - \widehat{r}_j|^3} - \frac{d\widehat{r}_i}{dt}, \tag{38}$$

where the hat denotes a unit vector ($|\widehat{r}_i(t)| \equiv 1$). The last term of Eq. (38) plays the role of an artificial viscosity to stabilize the system. At $t = 0$, $N_B$ charged particles are randomly distributed on the sphere, and they move around on the sphere according to Eq. (38). Since our purpose here is not to precisely trace the temporal evolution of the $N_B$-body system but to find the lowest potential configuration at $t \to \infty$, we can employ a relatively coarse time increment $\Delta t \sim 0.1$ in the simulations for $N_B \lesssim 100$. In this case, the iteration stops typically after a few hundred time steps at the solution when the variation of the potential energy $E_p$ between two succeeding steps is set to be less than $10^{-15}$, where $E_p$ is defined by

$$E_p = \frac{1}{2} \sum_{i=1}^{N_B} \sum_{j=1 (j \neq i)}^{N_B} \frac{1}{|\widehat{r}_i - \widehat{r}_j|}. \tag{39}$$

Fig. 7 shows the temporal evolution of the positions of twelve charged particles on a sphere at time steps, $t = 0, 20, 100$, and $200$, obtained from a simulation for M12. At $t = 0$, the twelve particles are put on the sphere clustered close to each other. The particles on the sphere are projected on the *xy*- (or equatorial-) plane normal to the view axis (*z*-axis), where the blue and red colors denote that they are on the north and south hemispheres, respectively. The concentric dashed curves stand for the projections of equi-latitude circles of 15°, 30°, 45°, 65°, and 75° measured from the pole axis. In the final configuration at $t = 200$, the twelve particles settle down at the face centers of the dodecahedron, as expected. Thereby ten particles, except for the two at the poles, all have polar angle $\theta = \arccos(1/\sqrt{5}) = 63.43°$ or $116.6°$, being positioned symmetrically on the same equi-latitude circles.

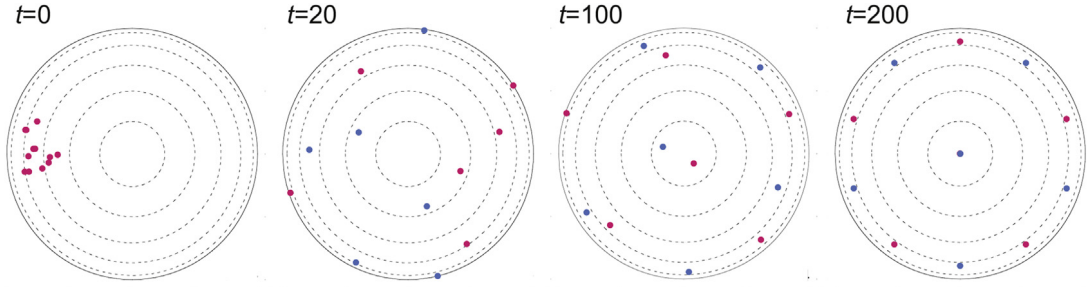

Fig. 7. Temporal evolution of the positions of twelve charged particles on a sphere at different time steps. The points are projected on the *xy*- (or equatorial-) plane normal to the view axis (*z*-axis), where the blue and red colors denote whether they are on the north or south hemisphere, respectively. The concentric dotted circles denote isocontour plane with polar angles of 15°, 30°, …, 75°. The final configuration at $t = 200$ coincides with that of the dodecahedron (referred to the inset in Fig. 8.)

Fig. 8 shows the temporal evolution of the relative potential energy for $N_B = 12$, defined by $\Delta E_p = E_p(t) - E_{p\infty}$, where $E_{p\infty} \equiv E_p(\infty)$. The solid curve corresponds to Fig. 7, in which the twelve face centers of the dodecahedron (see the 3D illustration) are reproduced at the end of calculation ($t \gtrsim 200$). The dashed curve corresponds to another case: the twelve particles are randomly distributed on the sphere at $t = 0$. The uppermost dotted curve has the same initial configuration as in Fig. 7, but with a different time increment $\Delta t = 0.6$. This simulation never converges to a stable configuration. Although the solid and dashed curves evolve on different trajectories in Fig. 8, both of them finally converge to the same configuration, i.e., the dodecahedron. As long as the number of particles is kept relatively small ($N_B \lesssim 30$), the final configurations obtained by the present method do not depend on their initial configurations at all. For larger numbers of charges, the final solution depends on the initial condition to give infinitesimally different final potential energies $E_p$. Such different solutions correspond to local minima in the potential field, though they generally provide very similar degree of illumination performance to each other.

We here again assume for simplicity that the corona layer, in which the laser rays are propagated with their energies being absorbed along the trajectories and finally scattered out of the target sphere, is thin enough and that all the incident laser rays are parallel to the beam axis. According to the separation of the single beam spectrum ($a_l$) and the pointing configuration ($\Omega_l$) as given in Sec. 2, the results obtained by the present self-organizing mechanism are not affected by the size or pattern of a laser beam, as long as they are assumed to be axisymmetric. The irradiation configuration of such a single beam is depicted in Fig. 9, where $r$ and $r_0$ are the distance of a ray measured from the beam axis and the target radius, respectively. Each ray is obliquely incident on the target surface with a pointing angle $\Psi = \arcsin(r/r_0)$ to the surface normal. As treated in Sec. 4, the corona plasma is approximated to have an exponentially decaying density profile along the radial direction. Then the absorption efficiency of a laser ray with a pointing angle $\Psi$ is computed [47] by $\eta_a(\Psi) = 1 - (1 - \eta_\perp)^{\cos^3\Psi}$. Furthermore, we employ a super-Gaussian intensity profile as a test profile of the single beam in Eq. (30): $I_L(r) = I_0 \exp[-(r/\alpha r_0)^\beta]$, where $I_0$ is the laser intensity on the beam axis, and $\alpha$ and $\beta$ are the control parameters of the beam shape. Since the target polar angle $\theta$ and the incident ray angle $\Psi$ coincide due to the parallel ray condition ($\theta = \Psi$), the resultant absorbed intensity distribution of a single beam is given as a function of $\theta$ only by

$$I_a(\theta) = I_0 \left[1 - (1 - \eta_\perp)^{\cos^3\theta}\right] \exp\left[-(\sin\theta/\alpha)^\beta\right]\cos\theta. \quad (40)$$

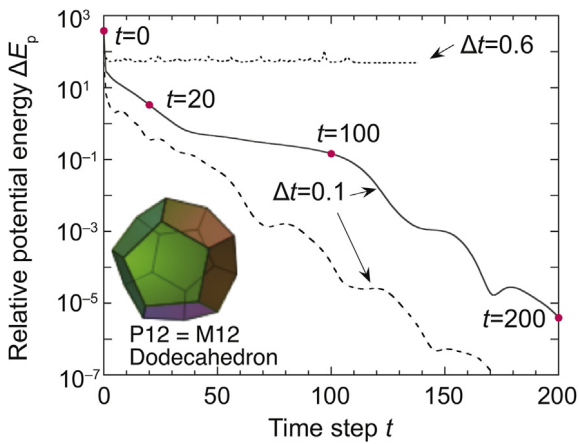

Fig. 8. Temporal evolution of $\Delta E_p$ for $N_B = 12$. The solid and dashed curves corresponds to the case of Fig. 1 and another with random distribution of particles at $t = 0$, respectively. Regardless of the initial configuration of particles, the algorithm always gives the same result, i.e., dodecahedron.

Fig. 10 shows the rms nonuniformity for various number of $N_B$ ranging from 10 up to 100 under the optimized configurations obtained by the self-organizing method; fixed parameters are $\eta_\perp = 90\%$, $\beta = 5$ (in Fig. 10(a)) and $\beta = 6$ (in Fig. 10(b)) and four different values of $\alpha$ for the super-Gaussian shape of an applied laser beam (referred to Eq. (30)). The overall behavior of the curves is similar to each other; after the quick drop for $N_B = 10-20$, $\sigma_{rms}$ monotonically decreases for $N_B = 20-60$. For the range of $60 \lesssim N_B \lesssim 100$, $\sigma_{rms}$ is nearly leveled off. Apart from these general behavior, two outstanding performances are found at $N_B = 48$ and $N_B = 72$, that are more conspicuous on Fig. 10(a). The eigenstructures of the two configurations, M48 and M72, are clarified below.

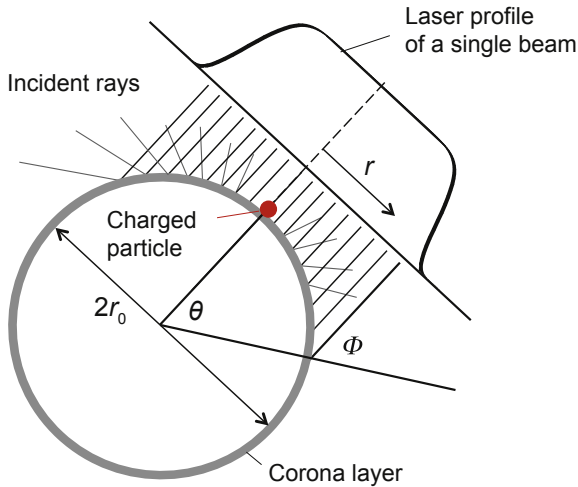

Fig. 9. Schematic view of applied laser and the target; $\theta$ is the polar angle at an observed point, and $\Phi$ is the incident angle of a laser. It is assumed that the corona layer is thin enough and the laser rays are incident on the spherical target surface parallel to the laser axis (in this case $\theta = \Phi$). In the self-organizing Coulomb system, a pseudo charged particle (red circle) is supposed to stand for the crossing point between a beam axis and the target surface.

Fig. 11 shows perspective views of some irradiation configurations; M24, M48, and M72 are obtained by the present method, while Ω24 and Ω60 are ones employed at Laboratory of Laser Energetics (LLE, University of Rochester). For M48 and M72 in Fig. 11, the solid and open circles denote that they are assigned to opposite hemispheres to each other. Laser beams are expected to irradiate the target with their beam axes passing through the vertices and the target center. Note that the three-dimensional (3D) views for M24, Ω24, and Ω60 show only the topological structures of the corresponding solids of the 13 Archimedean solids [36] (for a detailed description, for example, see Ref. [25]); the optimum beam positions obtained by the present method are slightly different from those 3D views.

Careful observation reveals that M24, M48 and M72 have very symmetric patterns. For example, in M48, equilateral triangles and squares are regularly located on the faces of a truncated octahedron. In the case of M72, twelve regular pentagons, each of which is composed of 6 points including the central points, are located on the dodecahedron faces, as can be seen in Fig. 11. Moreover, most of the optimized patterns obtained by the present method have no pair of points that are antipodes of each other (the twelve face centers of dodecahedron in M72 are rather exceptional). This is a pronounced advantage in view of optical system protection. Besides, the present algorithm provides the best irradiation configuration not only for even numbers but also for odd numbers of $N_B$.

The three configurations, M24, M48, and M72, have particularly symmetric structures such that they are expressed in a very compact way as follows: The left column of Table 3 gives the minimum data set for the exact spherical coordinates of M24 configurations, where only four pairs of coordinates $(\theta_i, \phi_i)$ for $i = 1, 2, 3, 4$ are given. Applying all the possible parametric combinations of the integers, $i$, $m$, and $l$, the coordinates of 24 (=4 × 2 × 3) beams of M24 measured in degrees are generated by

$$\text{M24} \begin{cases} \theta_{i+4m+8l} = (-1)^m \theta_i + 180m, \\ \phi_{i+4m+8l} = (-1)^m \phi_i + 95.6192m + 120l, \\ i = 1, 2, 3, 4; \quad m = 0, 1; \quad l = 0, 1, 2. \end{cases} \quad (41)$$

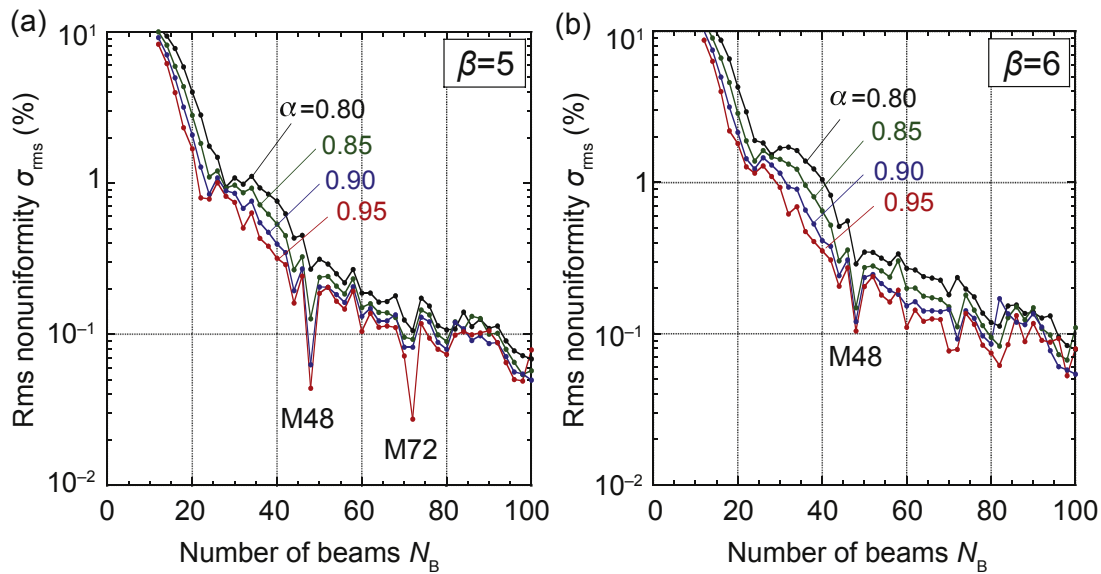

Fig. 10. The rms nonuniformity for various number of $N_B$. Their configurations are obtained by the self-organizing method. Fixed parameters are $\eta_\perp = 90\%$, (a) $\beta = 5$ and (b) $\beta = 6$ with four different values of $\alpha$, for the super-Gaussian shape of an applied laser beam (see Eq. (30)).

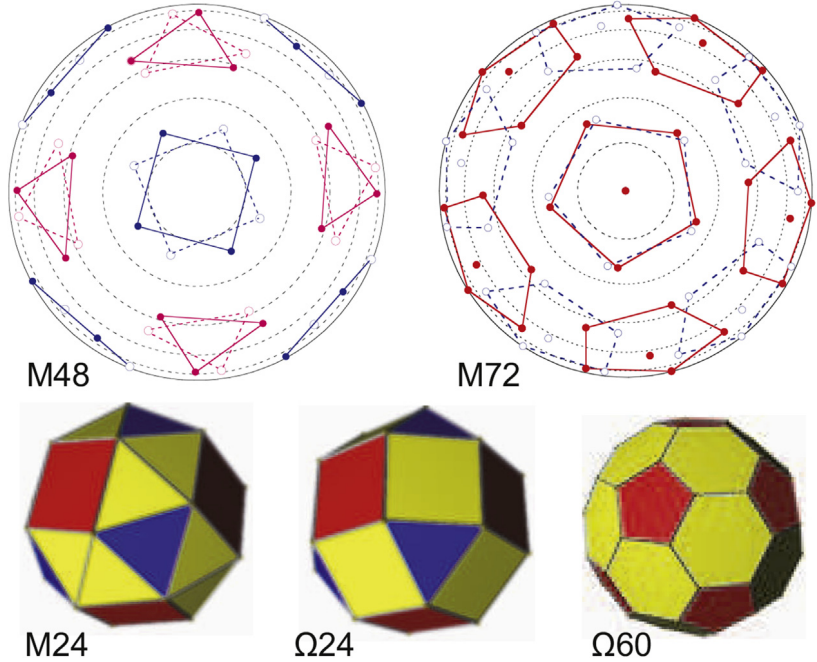

Fig. 11. Perspective views of some characteristic irradiation configurations. M24, M48, and M72 are obtained by the present self-organizing method, while Ω24 and Ω60 are ones of the actual laser systems at LLE. Laser beams are expected to irradiate the target with beam axes through the vertices of those solid bodies. The solid and dotted circles of M48 and M72 denote that they are on the opposite hemispheres to each other, and their exact spherical coordinates, $(\theta, \phi)$, are given by Eqs. (41)–(43) together with Table 3.

The integers, $m$ and $l$, perform the periodic projection of the basic coordinates in the table onto $\theta$ and $\phi$ coordinates, respectively. The four basic points in Table 3 form a square. In the same manner, the middle and right columns of Table 3 give the minimum data set for the exact spherical coordinates of M48 and M72, respectively, where only six and seven pairs of coordinates $(\theta_i, \phi_i)$ are given, and all the coordinates of 48 (=6 × 2 × 4) and 72 (=7 × 2 × 5 + 2) beams measured in degrees are generated according to the following formulas:

$$\text{M48} \begin{cases} \theta_{i+6m+12l} = (-1)^m \theta_i + 180m, \\ \phi_{i+6m+12l} = (-1)^m \phi_i + 37.2604m + 90l, \\ i = 1, 2, \ldots, 6; \quad m = 0, 1; \quad l = 0, 1, 2, 3, \end{cases} \quad (42)$$

$$\text{M72} \begin{cases} \theta_{i+7m+14l} = (-1)^m \theta_i + 180m, \\ \phi_{i+7m+14l} = (-1)^m \phi_i + 88.8328m + 72l, \\ i = 1, 2, \ldots, 7; \quad m = 0, 1; \quad l = 0, 1, 2, 3, 4, \\ (\theta_{71}, \phi_{71}) = (0, 0), \quad (\theta_{72}, \phi_{72}) = (180, 0). \end{cases} \quad (43)$$

Here it should be noted that 60- and 72-beam designs including Ω60 have the lowest dominant mode number,

Table 3
Minimum data set to generate the spherical coordinates of M24, M48, and M72 configurations (measured in degrees), all the coordinates of which are generated by using this table contents and Eqs. (41)–(43).

| $i$ | M24 | | M48 | | M72 | |
|---|---|---|---|---|---|---|
| | $\theta_i$ | $\phi_i$ | $\theta_i$ | $\phi_i$ | $\theta_i$ | $\phi_i$ |
| 1 | 26.24800 | 0 | 21.24302 | 0 | 24.49171 | 84.24238 |
| 2 | 52.56226 | 55.84264 | 43.64296 | 43.69981 | 41.30650 | 48.91327 |
| 3 | 67.23147 | 345.7672 | 51.30717 | 86.62623 | 49.87536 | 87.02248 |
| 4 | 84.37483 | 25.62150 | 69.91959 | 25.70934 | 63.43495 | 62.41641 |
| 5 | | | 72.99624 | 59.45306 | 67.52800 | 35.82484 |
| 6 | | | 83.35323 | 88.40802 | 77.90530 | 83.44693 |
| 7 | | | | | 87.27933 | 56.58353 |

$n_d = 10$, in common as long as they are based on one of the Platonic solids [25], i.e., dodecahedron. In contrast, the present method gives $n_d = 12$ for M48, $n_d = 14$ for M60, and $n_d = 15$ for M72, which are significant advantage of the new configurations.

Fig. 12 shows the irradiation nonuniformity $\sigma_{\text{rms}}$ defined by Eqs. (15)–(17) versus the system imperfection $\sigma_{\text{sys}}$ for different irradiation configurations. The employed parameters are $\alpha = 0.9$ and $\beta = 6$ for the beam profile, from which $\sigma_{\text{rms}}^{\text{single}} = 2.5$ is computed, and the absorption is set to be $\eta_\perp = 95\%$ just as an example. The spectrum for the absorbed single beam factor shows approximately an exponential decrease with the mode number such that $a_l / \sqrt{2l+1} \approx 4.0 \exp(-0.48\, l)$. There are two important messages that can be read from Fig. 12: First, it is crucial to optimize the irradiation system coherently with the beam pattern. In the case of Fig. 12, for instance, M48 shows comparable irradiation uniformity to those of M60 and M72. Second, even if the irradiation system can achieve a super-high irradiation uniformity under idealized circumstance, i.e., $\sigma_{rms}^0 \ll 1$ at $\sigma_{sys} \to 0$, this would be useless if the system imperfection is beyond the critical value $\sigma_{sys}^{crt} \approx (\sigma_{rms}^0 / \sigma_{rms}^{single}) \sqrt{N_B}$. For $\sigma_{sys} > \sigma_{sys}^{crt}$, the rms nonuniformity can be roughly estimated by $\sigma_{rms} \approx 2\sigma_{sys}/\sqrt{N_B}$. This can be easily confirmed from Fig. 12, where the four curves corresponding to M48, M60, M72, and $\Omega 60$ are almost merged for $\sigma_{sys} \gtrsim 3\%$ despite having clean rms nonuniformities $\sigma_{rms}^0$ distributed at different levels from each other for $\sigma_{sys} \ll 2\% - 3\%$. Note that high-mode non-uniformities attributed to speckle patterns are not taken into account here, the inclusion of which will smear to some degree and lead to distinctive differences among the curves depicted in this figure. Also note that the two different configurations, M48 and P32, are numerically compared in more detailed manner in Ref. [20].

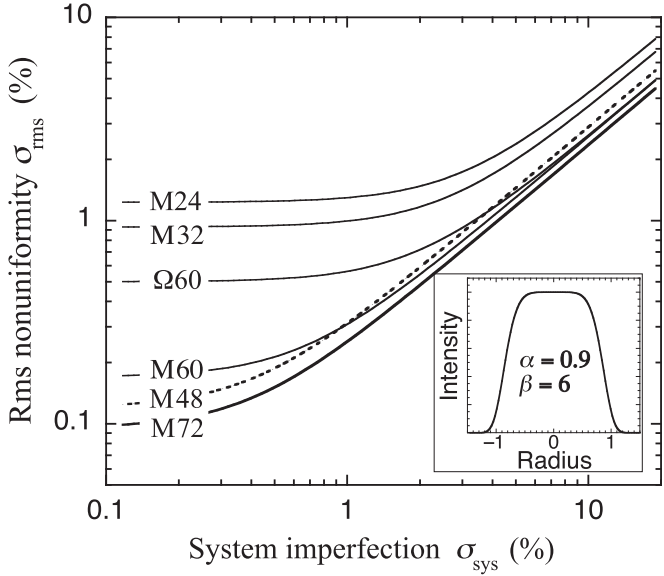

Fig. 12. The rms irradiation nonuniformity versus the system imperfection. The inset shows the applied laser intensity profile in a super-Gaussian shape; fixed parameters are $\alpha = 0.9$, $\beta = 6$, and $\eta_\perp = 95\%$ (see Eqs. (29) and (30)). An important message: Unless system imperfections cannot be effectively suppressed below a few percent, any benefits of an optimized beam configuration are overwhelmed.

Fig. 13 plots isocontour maps of $\sigma_{rms}^0$ and the integrated absorption efficiency,

$$\overline{\eta}_a = \frac{\int_0^1 \xi \exp\left[-(\xi/\alpha)^\beta\right] I_a(\theta) d\xi}{I_0 \int_0^\infty \xi \exp\left[-(\xi/\alpha)^\beta\right] d\xi}, \quad (44)$$

where $I_a(\theta)$ is given by Eq. (40) with $\theta = \arccos\sqrt{1-\xi^2}$, showing their dependence on the applied laser intensity profile, Eq. (30), as a function of $\alpha$ and $\beta$, for the three configurations, $\Omega 60$, M48, and M60; $\eta_\perp$ is fixed at 95% to draw the curves of $\overline{\eta}_a$. All these configurations have working windows, where relatively high efficiency and uniformity are achieved simultaneously. Such performance can be assessed by comparing the area that satisfies both $\overline{\eta}_a \geq 60\%$ and $\sigma_{rms} \leq 0.4\%$, for instance. Here again it should be noted that Fig. 13 is obtained under several specific numbers and assumptions and therefore they may substantially differ depending on individual and practical situations. Nevertheless, a number of such parameter surveys have shown that, in spite of the smaller number of beams, the illumination performance of M48 is comparable enough to that of $\Omega 60$ (also compare with Fig. 1). Furthermore, moderate slopes on the contour map in Fig. 13 lead to more stable illumination performance with variation of the laser intensity profile. From those evaluations, M60 is generally expected to provide even better illumination performance than M48 or $\Omega 60$.

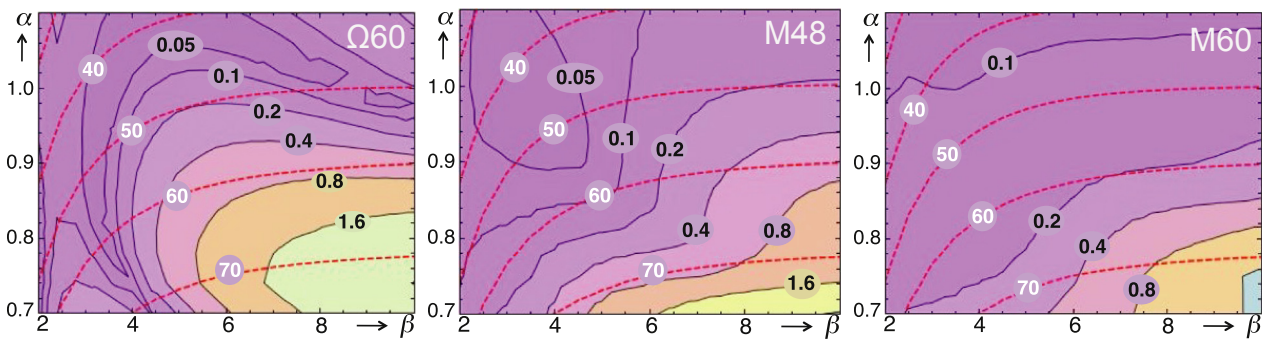

Fig. 13. Isocontour maps of the clean rms nonuniformity $\sigma_{rms}^0$ (purple curves) and the integrated absorption efficiency $\overline{\eta}_a$ (red dashed curves) showing their dependence on the applied laser intensity profile (see Eq. (30)) as a function of $\alpha$ and $\beta$. The numbers are all in units of %, and $\eta_\perp$ is fixed at 95% to draw the curves for $\overline{\eta}_a$. Three different configurations, $\Omega 60$, M48, and M60, correspond to Omega-Upgrade, and optimized 48- and 60-beam configurations obtained by the self-organizing Coulomb method, respectively.

## 6. Summary

The purpose of optimizing laser illumination configuration is to suppress lowest mode nonuniformities for $l \lesssim 10-20$ with laser beams of $N_B \lesssim 50-100$ as much as possible. We have reviewed the schemes to optimize configuration of laser illumination for direct-drive. Under the assumption that a single laser beam has an axisymmetric absorption pattern, the illumination performance has been discussed in terms of the number of laser beams and system imperfection as well as the absorbed laser profile of a single beam. Based on the analytical model, the rms nonuniformity is factorized into two components, the single beam factor, $a_l$, and the geometrical factor, $G_l$.

For such a system in which all the beam axes go through the target center, the illumination system turns out to be optimized in terms of the geometrical factor alone. Besides, we have proposed a new PDD scheme, with which lower modes nonuniformity can be effectively suppressed by appropriately choosing the stand-off distances of the beam axes under the optimized combination of the incident angle of beam axes and assigned energies of the illuminating cones. The off-center illumination scheme is expected to improve the irradiation uniformity of such laser systems as NIF or LMJ that are originally constructed for indirect-drive. The practical application of the present PDD method is still premature and valid only for limited cases, since the dynamics and 3D features of laser absorption are not taken into account. More detailed analysis will be reported somewhere else.

A simple numerical model, which is based on a self-organizing mechanism of charged particles on a sphere, is presented to give the best configuration for an arbitrary number of laser beams. It has turned out, for example, that one of such configurations, M48 or M60, shows comparable or even better illumination performance than any other existing configurations.

## Acknowledgements

This work was supported by the Japan Society for the Promotion of Science (JSPS).